\documentclass[pra,superscriptaddress,twocolumn,showpacs,floatfix]{revtex4}
\usepackage{graphicx}
\newcommand{\Journal}[4]{#1 {\bf #2}, #3 (#4)}
\newcommand{\PR}{Phys. Rev.}
\newcommand{\PRL}{Phys. Rev. Lett.}
\newcommand{\PRA}{Phys. Rev. A}
\newcommand{\JMP}{J. Math. Phys.}
\newcommand{\Science}{Science}
\newcommand{\PLA}{Phys. Lett. A}
\def\la{\langle} 
\def\ra{\rangle}
\def\om{\omega}

\newcommand{\beq}{\begin{equation}}
\newcommand{\eeq}{\end{equation}}
\newcommand{\beqa}{\begin{eqnarray}}
\newcommand{\eeqa}{\end{eqnarray}}
\begin{document}
\title {Stability of spinor Fermi gases in tight waveguides}
\author{A. del Campo}
\email{qfbdeeca@ehu.es} \affiliation{Departamento de Qu\'\i mica-F\'\i sica, 
Universidad del Pa\'\i s Vasco, Apdo. 644, 48080 Bilbao, Spain}
\author{J. G. Muga}
\email{jg.muga@ehu.es} \affiliation{Departamento de Qu\'\i mica-F\'\i sica, 
Universidad del Pa\'\i s Vasco, Apdo. 644, 48080 Bilbao, Spain}
\author{M. D. Girardeau}
\email{girardeau@optics.arizona.edu}
\affiliation{College of Optical Sciences, University of Arizona,
Tucson, AZ 85721, USA}
\date{\today}
\begin{abstract}
The two and three-body correlation functions of the ground state of an 
optically trapped ultracold spin-$\frac{1}{2}$ Fermi gas (SFG) in a tight 
waveguide (1D regime) are calculated in the plane of even and odd-wave
coupling constants, assuming a 1D attractive zero-range odd-wave interaction
induced by a 3D p-wave Feshbach resonance, as well as the usual repulsive
zero-range even-wave interaction stemming from 3D s-wave scattering.
The calculations are based on the exact mapping from the SFG to a
``Lieb-Liniger-Heisenberg'' model with delta-function
repulsions depending on isotropic Heisenberg spin-spin interactions,
and indicate that the SFG should be stable against three-body recombination
in a large region of the coupling constant plane encompassing parts of both the
ferromagnetic and antiferromagnetic phases. However, the limiting case of the fermionic 
Tonks-Girardeau gas (FTG), a spin-aligned 1D Fermi gas with infinitely attractive 
p-wave interactions, is unstable in this sense. 
Effects due to the dipolar
interaction and a Zeeman term due to a resonance-generating magnetic field
do not lead to shrinkage of the region of stability of the SFG.
\end{abstract}
\maketitle

\section{Introduction}

At the low densities of ultracold bosonic atomic vapors, the interatomic 
interactions are accurately parametrized by the 3D s-wave scattering length
$a_s$. It was shown by Olshanii \cite{Ols98,BerMooOls03,PetShlWal00} that when 
such a vapor
is confined in a de Broglie wave guide with transverse trapping so tight and 
temperature so low that the transverse vibrational excitation quantum 
$\hbar\omega_{\perp}$ is larger than available longitudinal zero point and 
thermal energies, the effective dynamics becomes one-dimensional (1D)
and accurately described by a 1D Hamiltonian with delta-function interactions
$g_{1D}\delta(x_j-x_{\ell})$, where $x_j$ and $x_{\ell}$ are 1D longitudinal
position variables. This is a famous integrable model, the Lieb-Liniger (LL) 
model, exactly solved in 1963 by a Bethe ansatz method \cite{LieLin63}.
The value of $g_{1D}$ is a known function 
\cite{Ols98,BerMooOls03}
of $a_s$ exhibiting a confinement-induced 1D Feshbach resonance (CIR), such 
that the value of the 1D scattering length $a_{1D}$ can be tuned from $-\infty$
to $+\infty$ by shifting the position of a 3D Feshbach scattering resonance
(hence the value of $a_s$) via an external magnetic field \cite{Rob01}.
Such CIRs also occur in spin-aligned Fermi gases in tight waveguides 
\cite{GraBlu04}. Near the CIRs the ground states have
strong short-range correlations not representable by effective field
theories, and such systems have become the subject of extensive 
theoretical and experimental investigations \cite{Tol04,Mor04,Par04Kin04,Kin05,Kin06}, 
particularly since the 
recent experimental realization \cite{Par04Kin04} of the 1D gas of impenetrable
point bosons ($g_{1D}\to\infty$ limit of the LL model), solved
exactly in 1960 \cite{Gir60,Gir65} by Fermi-Bose (FB) mapping to the ideal
Fermi gas, and now known as the Tonks-Girardeau (TG) gas. 

The ``fermionic Tonks-Girardeau'' (FTG) gas is the ``mirror image'' of the
TG gas; instead of 1D bosons with infinitely strong zero-range repulsions,
one has 1D spin-aligned fermions with infinitely strongly attractive zero-range
odd-wave interactions. The exact ground state of this system was determined 
recently \cite{GraBlu04,GirOls03,GirOls04-1,GirNguOls04,GirMinGir06} by FB 
mapping to the ideal Bose gas; this system
models a magnetically trapped ultracold gas of spin-$\frac{1}{2}$ fermionic 
atoms with infinitely strong odd-wave interactions induced by a p-wave Feshbach
resonance. More generally, by considering a spin-aligned Fermi gas with
a very strong but finite interaction modelled by a very deep and narrow
square well interaction of depth $V_0$ and width $2x_0$ and carrying out
the zero-range limit such that $V_{0}x_0^2$ is held constant as $V_0\to\infty$
and $x_0\to 0$, one obtains a model with a finite and negative 1D scattering
length $a_{1D}$ determined by the value of $V_{0}x_0^2$, which is exactly
soluble \cite{GraBlu04,GirOls03,GirOls04-1,GirNguOls04} by FB mapping to the LL 
1D Bose gas, and models a spin-aligned Fermi gas in a tight waveguide in the
neighborhood of a 3D p-wave Feshbach resonance with an associated 1D 
odd-wave CIR.

If optically trapped instead, such a spinor Fermi gas (SFG) has 
richer properties since its spins are unconstrained, and there are
both spatial even-wave interactions
associated with spin singlet scattering and odd-wave interactions associated
with spin triplet scattering. One can vary the ratio of the two 
coupling constants by Feshbach resonance tuning of the odd-wave one,
leading to a rich phase diagram of ground state spin, with both ferromagnetic
and antiferromagnetic phases \cite{GirNguOls04,Note1,Gir06b}.

In \cite{Gir06b} the ground and low excited states of the SFG were calculated 
via a generalized FB mapping to a 1D Lieb-Liniger-Heisenberg (LLH) model 
of particles with delta-function repulsions depending on Heisenberg 
spin-spin interactions, allowing exact determination of the ground state
energy in the ferromagnetic phase in terms of the LL Bethe 
ansatz solution \cite{LieLin63}, and a variational calculation in the
antiferromagnetic phase combining the exact Bethe ansatz solutions of the LL 
model and the 1D Heisenberg antiferromagnet. For experiments on such
a system the local correlation functions are very important. The three-body
correlation function $g_3$ determines the rate of three-body recombination
and hence the stability of the system against three-body decay
\cite{Tol04}, and the two-body correlation function $g_2$ determines the
rate of photoassociation to excited diatomic molecules, a process which
has been used recently to experimentally confirm \cite{Kin05}
fermionization of bosons in accord with the FB mapping \cite{Gir60,Gir65}.
The purpose of the present paper is to determine these correlation functions and thus, 
the stability of the SFG gas in terms of the coupling constants 
governing the odd and even-wave interactions.

\section{SFG and LLH models} 

The SFG Hamiltonian in LLH units $\hbar=2m=1$
\cite{LieLin63} is \cite{Gir06b}
\begin{equation}\label{Fermi Hamiltonian}
\hat{H}_{\text{SFG}}=-\sum_{j=1}^{N}\partial_{x_j}^{2}
+\sum_{1\le j<\ell\le N}[g_{1D}^{e}\delta(x_{j\ell})\hat{P}_{j\ell}^s
+v_{1D}^{\text{o}}(x_{j\ell})\hat{P}_{j\ell}^t]\ .
\end{equation}
Here $x_{j\ell}=x_j-x_{\ell}$, 
$\hat{P}_{j\ell}^s=\frac{1}{4}-\hat{\mathbf{S}}_j\cdot\hat{\mathbf{S}}_{\ell}$ 
and 
$\hat{P}_{j\ell}^t=\frac{3}{4}+\hat{\mathbf{S}}_j\cdot\hat{\mathbf{S}}_{\ell}$
are the projectors onto the subspaces of singlet and triplet functions of the 
spin arguments $(\sigma_j,\sigma_\ell)$ for fixed values of all other 
arguments, and $v_{1D}^{\text{o}}$ is a strong, attractive, zero-range, 
odd-wave interaction (1D analog of 3D p-wave interaction)
which is the zero-range limit of a deep and narrow square well of depth 
$V_0$ and width $2x_0$, where the zero-range limit $x_0\to 0+$ and 
$V_0\to\infty$ is taken such that 
$\sqrt{V_0/2}=(\pi/2x_{0})[1+(2/\pi)^2 (x_{0}/a_{1D}^{o})]$, thus generating
a relative wave-function with a contact discontinuity in the zero-range limit 
\cite{CheShi98,GirOls03,GirOls04-1,GirNguOls04} and satisfying the contact 
condition
\beqa
\psi_{F}(x_{j\ell}=0+)=-\psi_{F}(x_{j\ell}=0-)
= -a_{1D}^{o}\psi_{F}^{'}(x_{j\ell}=0\pm)\nonumber, 
\eeqa
where $a_{1D}^{o}<0$ is the 1D odd-wave scattering length and the prime denotes
differentiation.
The even-wave 1D coupling constant $g_{1D}^{e}$ in (\ref{Fermi Hamiltonian})
is related to the even-wave scattering length $a_{1D}^{e}<0$ derived 
\cite{Ols98} from 3D s-wave scattering by $g_{1D}^{e}=-4/a_{1D}^e$ 
and the even-wave contact condition is the usual LL one \cite{LieLin63} 
\beqa
\psi_{F}'(x_{j\ell}=0+)
=-\psi_{F}'(x_{j\ell}=0-)=-\frac{\psi_{F}(x_{j\ell}=0\pm)}{a_{1D}^{e}}.
\nonumber
\eeqa
Assume that the system is contained in a ring trap of circumference $L$, with 
periodic boundary conditions of periodicity length $L$, the ring 
circumference. The eigenfunctions of $\hat{H}_{\text{SFG}}$ 
can be mapped to those of the LLH model by multiplying the part of the SFG wave
function 
which is odd in $x_{j\ell}$ by 
$\epsilon(x_{j\ell})$, where $\epsilon(x)=+1\ (-1)$ for $x>0\ (x<0)$, and 
$\epsilon(0)=0$, while leaving the even part unchanged \cite{Gir06b}; 
this generalizes the original FB mapping of \cite{Gir60} to the spin-dependent
case here, by converting the odd-wave contact discontinuities as $x_{j\ell}\to 0$ into contact cusps of the usual LL form \cite{LieLin63}. Moreover, 
the complete energy spectrum is unchanged in terms of a transformed
LLH Hamiltonian, with usual LL delta function interactions supplemented
by spin-spin interactions of the usual Heisenberg form arising from the
singlet and triplet spin projectors $\hat{P}_{j\ell}^s$ and
$\hat{P}_{j\ell}^t$. One finds \cite{Gir06b}
\begin{flushleft}
\begin{equation}\label{LLH Hamiltonian}
\hat{H}_{\text{LLH}}\!=\!-\sum_{j=1}^{N}\partial_{x_j}^{2}
\!+\!\!\!\!\sum_{1\le j<\ell\le N}\!\bigg[\frac{3c_o+c_e}{2}
+2(c_o-c_e)\hat{\mathbf{S}}_j\cdot\hat{\mathbf{S}}_{\ell}\bigg]\delta(x_{j\ell})
\end{equation}
\end{flushleft}
where the term in brackets plays the role of $2c$ in the LL model.
The coupling constants are related to the odd and even-wave SFG
scattering lengths $a_{1D}^o$ and $a_{1D}^e$ by
\beqa
c_o=\frac{2}{|a_{1D}^o|},\qquad c_e=\frac{1}{2}g_{1D}^e=\frac{2}{|a_{1D}^e|}.
\eeqa 
In a real optically trapped ultracold alkali metal vapor there are also 
dipolar interactions between the valence electron spins as well as their 
Zeeman interactions with an external magnetic field required to generate
the p-wave Feshbach resonance. The experimental results of Ticknor 
{\it et al.} for $^{40}$K \cite{Tik04} show that the dipolar interaction 
causes a very small ($\sim 1\%$) shift and splitting of the resonance,
thus shifting the 1D CIR and hence the value of $c_o$. However, $c_o$ is an
adjustable parameter in (\ref{LLH Hamiltonian}), so we shall not include 
the dipolar interaction explicitly. The effect of the Zeeman interaction
is more complicated and will be considered after we have evaluated the 
correlation functions of the ground state of (\ref{LLH Hamiltonian}).

The ground state of (\ref{LLH Hamiltonian}) and hence of 
(\ref{Fermi Hamiltonian}) is ferromagnetic with total spin $S=N/2$ for $c_o<c_e$, 
and antiferromagnetic with $S=0$ for $c_o>c_e$ \cite{GirNguOls04,Gir06b}. The 
ferromagnetic ground state is of space-spin product form
$\psi_{\text{space}}\psi_{\text{spin}}$ and is a simultaneous eigenstate
of the spin-spin interaction operators 
$\hat{\mathbf{S}}_j\cdot\hat{\mathbf{S}}_{\ell}$ with eigenvalue $\frac{1}{4}$
for each, reducing the Hamiltonian (\ref{LLH Hamiltonian}) to the usual LL form
with interactions $2c_o\delta(x_{j\ell})$ independent of $c_e$ and with the 
spatial ground state $\psi_{\text{space}}$ given exactly for all $c_o$ by the LL 
Bethe ansatz \cite{LieLin63}. The antiferromagnetic ground state is not known
exactly, but a natural variational approximation is given by a space-spin
product state $\psi_{\text{space}}\psi_{\text{spin}}$ with $\psi_{\text{spin}}$
the exact Bethe ansatz ground state of the antiferromagnetic Heisenberg
model. Ordering the particle coordinates $x_1<x_2<\cdots<x_N$ and replacing
$\hat{\mathbf{S}}_j\cdot\hat{\mathbf{S}}_{j+1}$ in (\ref{LLH Hamiltonian})
by its expectation value $\frac{1}{4}-\ln 2$ in the antiferromagnetic 
Heisenberg ground state \cite{Hul38} again reduces the spatial Hamiltonian to 
LL form, but now with the coupling constant $c$ of LL repaced by 
$c=c_o(1-\ln 2)+c_e\ln 2$ \cite{Gir06b}.

\section{Two-body correlations} 

\begin{figure}
\includegraphics[width=7.5cm,angle=0]{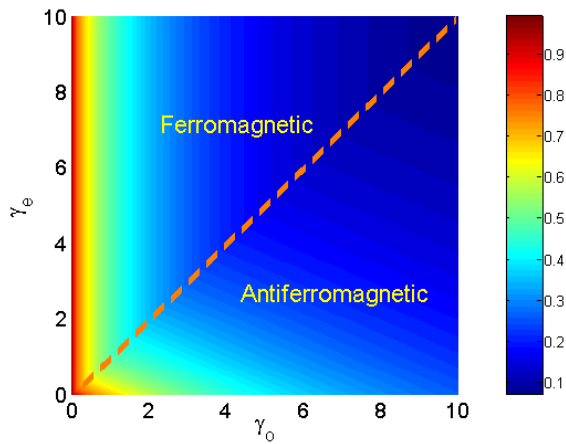} 
  \caption{
Plot of the two-body correlation function $g_2(\gamma_o,\gamma_e)/n^2$
of the SFG in the $\gamma_o$, $\gamma_e$-plane where the different phases are 
identified. The quantum phase transition lies along the diagonal dashed 
line $\gamma_o=\gamma_e$.  
}  \label{fig1}
\end{figure}

The two-body correlation function of the SFG, which determines the rate of photoassociation 
to excited diatomic molecules \cite{Kin05}, is equal to that of the LL model, 
$g_2(\gamma)=\la\hat{\Psi}^{\dag}(x)^2\hat{\Psi}(x)^2\ra_0$, 
where
$\gamma=c/n$ is the dimensionless coupling constant, $n=N/L$ is the 1D density,
$\hat{\Psi}^{\dag}$and $\hat{\Psi}$ are the LL boson creation and 
annihilation operators, and $\la\cdots\ra_0$ is the ground-state expectation 
value. It can be evaluated exactly in the thermodynamic limit using the 
Hellmann-Feynman theorem \cite{GanShl03,Kher03}: 
$g_2(\gamma)/n^2=de(\gamma)/d\gamma$ where $e$ is the scaled
ground state energy \cite{LieLin63} $e(\gamma)=n^{-2}(E_0/N)$ 
numerically calculated in \cite{Olshanii}.
For the SFG we shall use
\beqa
\gamma=\gamma_o
\eeqa
in the ferromagnetic phase and
\beqa
\gamma=\gamma_o(1-\ln 2)+\gamma_e\ln 2
\eeqa
 in the antiferromagnetic phase with 
$\gamma_o=c_o/n=2/(n|a_{1D}^o|)$ and $\gamma_e=c_e/n=2/(n|a_{1D}^e|)$. The results are
shown in Fig. \ref{fig1}, where the two-body correlation function is shown to reach 
values near unity in both 
ferromagnetic and antiferromagnetic phases when the p-wave interactions 
are highly attractive, which corresponds to small values of $\gamma_o$. 
Note that this concerns the limiting case of the FTG gas, 
defined as a spin-aligned 1D Fermi gas (hence, not subjected to the s-wave 
pseudopotential) 
with infinitely attractive odd-wave interactions ($\gamma_o=0$).
Indeed, for $\gamma\ll 1$, $g_2(\gamma)/n^2=1-2\sqrt{\gamma}/\pi$ whereas for $\gamma\gg1$, $g_2(\gamma)/n^2=4\pi^2/3\gamma^2$ \cite{GanShl03}.  In the TG 
limit of the LLH model, $\gamma\rightarrow \infty$, 
all local correlation functions vanish.

\section{Three-body correlations} 

The rate coefficient of three-body losses is 
proportional to the three-body correlation function 
$g_3(\gamma)=\la\hat{\Psi}^{\dag}(x)^3\hat{\Psi}(x)^3\ra_0$
as demonstrated experimentally \cite{Tol04}. The exact expression for 
$g_3(\gamma)$ was recently derived using an integrable lattice model, 
the $q$-boson hopping model, which 
reduces to the LL model in the thermodynamic limit 
\cite{CheSmiZvo06,CheSmiZvo06b},
\beqa
\frac{g_3(\gamma)}{n^3}&=&
\frac{3}{2\gamma}\epsilon_4'-\frac{5\epsilon_4}{\gamma^2}
+\left(1+\frac{\gamma}{2}\right)\epsilon_2'
\nonumber\\
& & -2\frac{\epsilon_2}{\gamma}-
\frac{3\epsilon_2\epsilon_2'}{\gamma}+\frac{9\epsilon_2^2}{\gamma^2},
\label{g3result}
\eeqa
where 
\beq
\epsilon_m=\left(\frac{\gamma}{\alpha}\right)^{m+1}
\int_{-1}^{1}{\rm d}k\,k^m\sigma(k) \qquad (m=2,4)
\eeq
are the moments of the quasi-momentum distribution \cite{Note2} which obeys the
usual LL linear integral equation \cite{LieLin63}, 
\beqa
\sigma(k)-\frac1{2\pi}\int_{-1}^1{\rm d}q\,\frac{2\alpha\sigma(q)}
{\alpha^2+(k-q)^2}=\frac1{2\pi},
\eeqa
with 
$\alpha=\gamma\int_{-1}^1{\rm d}k\,\sigma(k)$, and primes denote derivative with respect to $\gamma$.
Approximate expressions for $g_3(\gamma)$ with a relative error smaller than 
$2\times 10^{-3}$ were provided for $\gamma\in[0,30]$\cite{CheSmiZvo06}.
%
\begin{figure}
\includegraphics[width=7.5cm,angle=0]{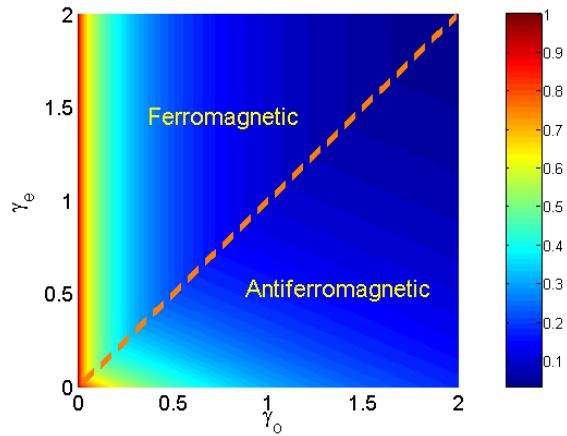} 
  \caption{Plot of the three-body correlation function 
$g_3(\gamma_o,\gamma_e)/n^3$ in the $\gamma_o$, $\gamma_e$-plane. 
} \label{fig2}
\end{figure}
%
Note that such results cover both asymptotics.
For $\gamma\ll 1$ the Bogoliubov result holds, 
$g_3(\gamma)/n^3\simeq 1-6\sqrt{\gamma}/\pi$ which in particular, 
points out the instability of the FTG gas with respect to three-body losses.
As $\gamma\rightarrow\infty$, 
$g_3(\gamma)/n^3\simeq 16\pi^6/15\gamma^6$.
Such limiting cases are known for the correlation functions of all orders 
\cite{GanShl03b}.
Fig. \ref{fig2} shows the decay of $g_3$ as a function of both $\gamma_o$ 
and $\gamma_e$ in both phases. Decay processes due to three-body collisions make 
the gas unstable for high attractive interactions but become 
negligible as soon as $\gamma\gtrsim 1$.

\section{Effect of the Zeeman interaction} 
%
\begin{figure}
\includegraphics[width=4.25cm,angle=0]{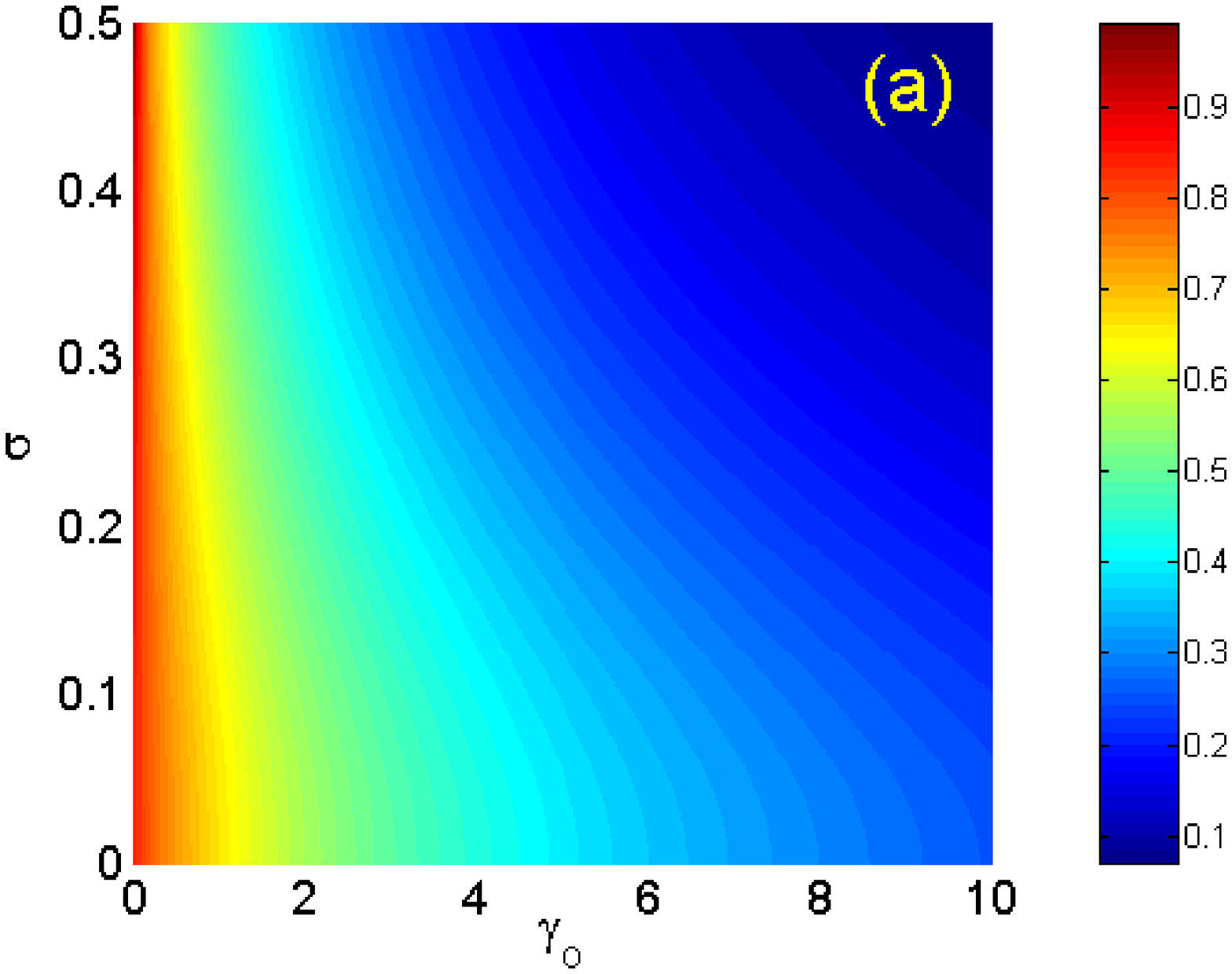}
\includegraphics[width=4.25cm,angle=0]{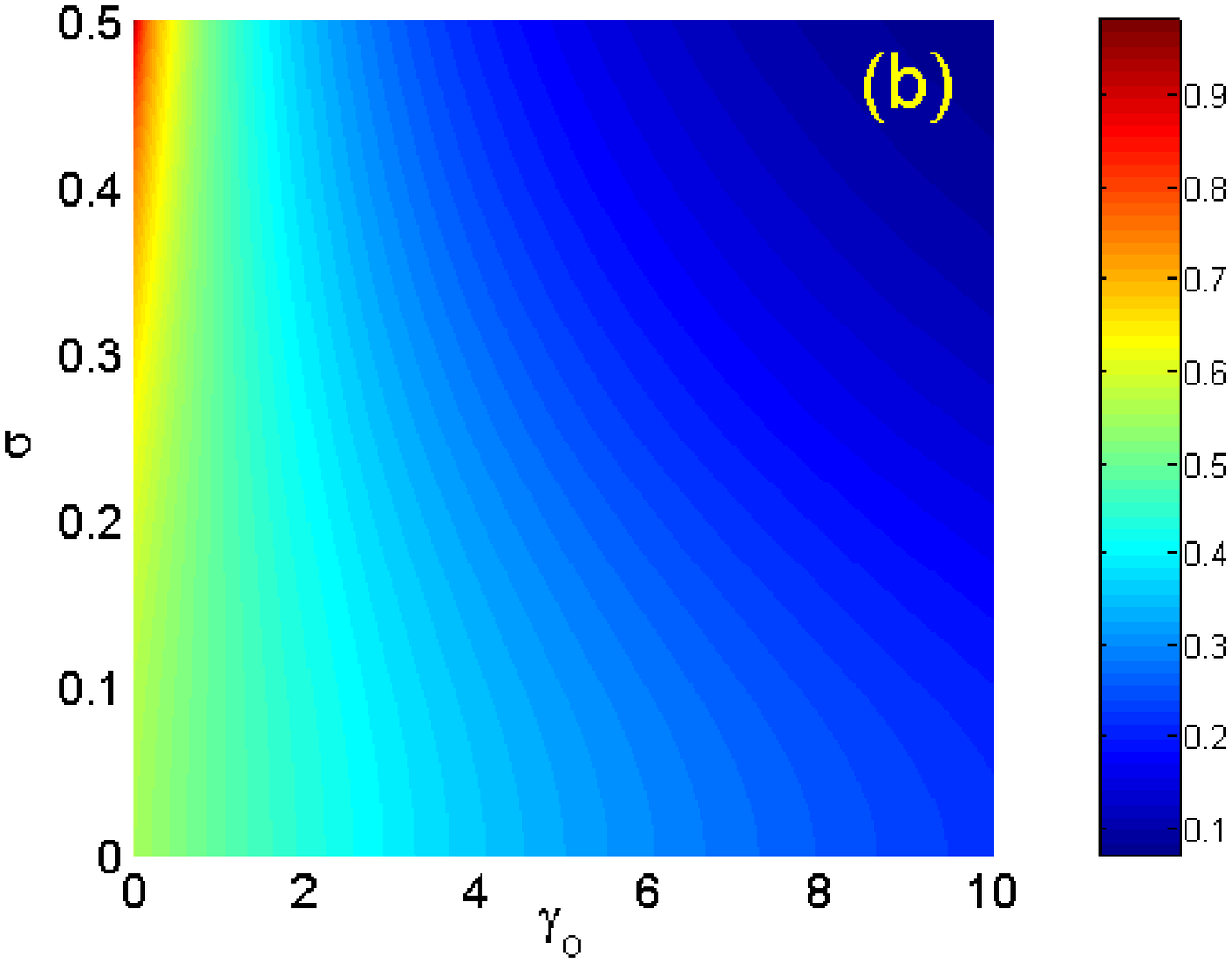} 
  \caption{Plot of the two-body correlation function 
$g_2(\gamma_o,\sigma;\gamma_e)/n^2$ in the $\gamma_o$, $\sigma$-plane with $\gamma_e$ 
equal to $0.1$ (a) and 1 (b).
}
  \label{fig3}
\end{figure}
%

%
\begin{figure}
\includegraphics[width=4.25cm,angle=0]{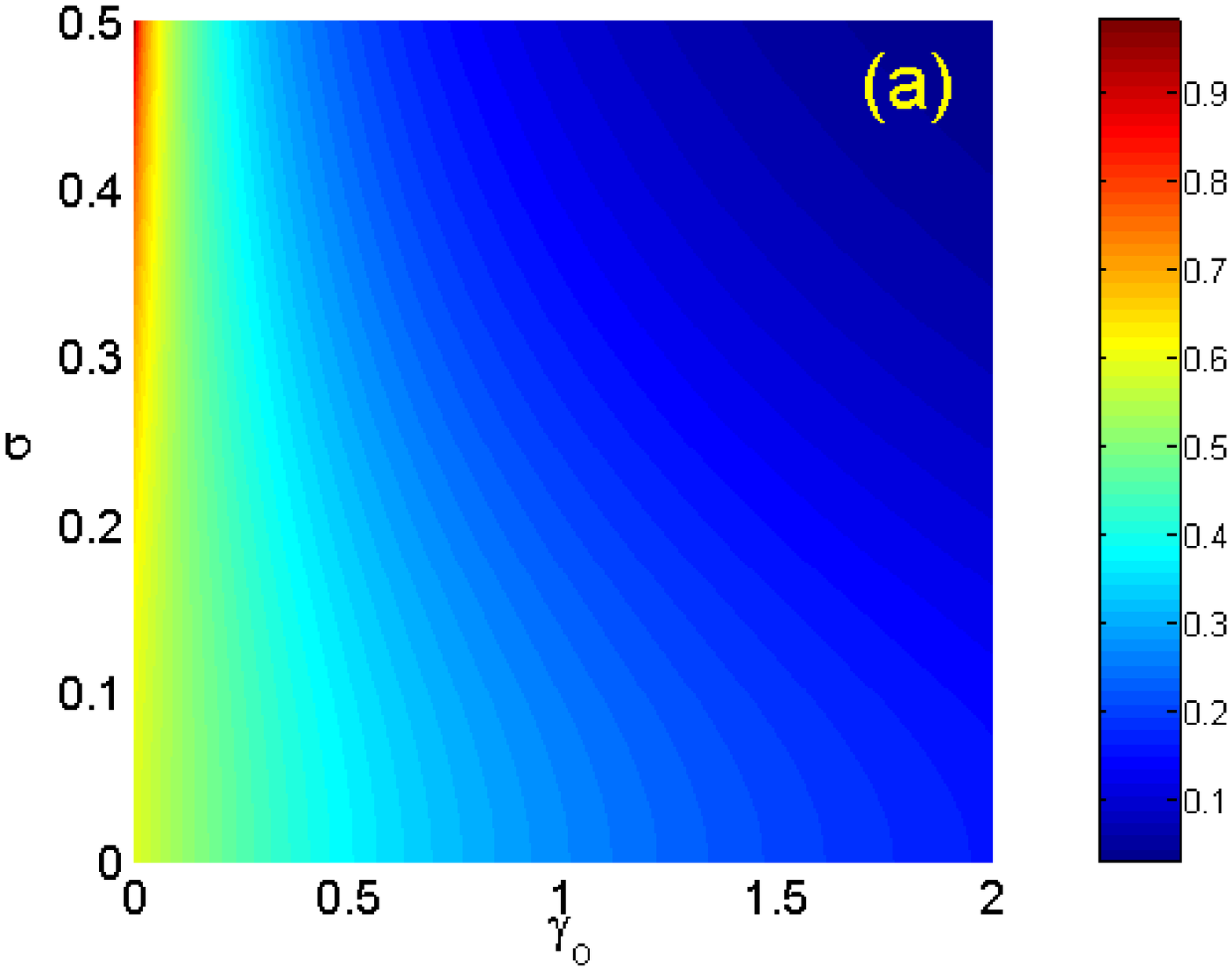}
\includegraphics[width=4.25cm,angle=0]{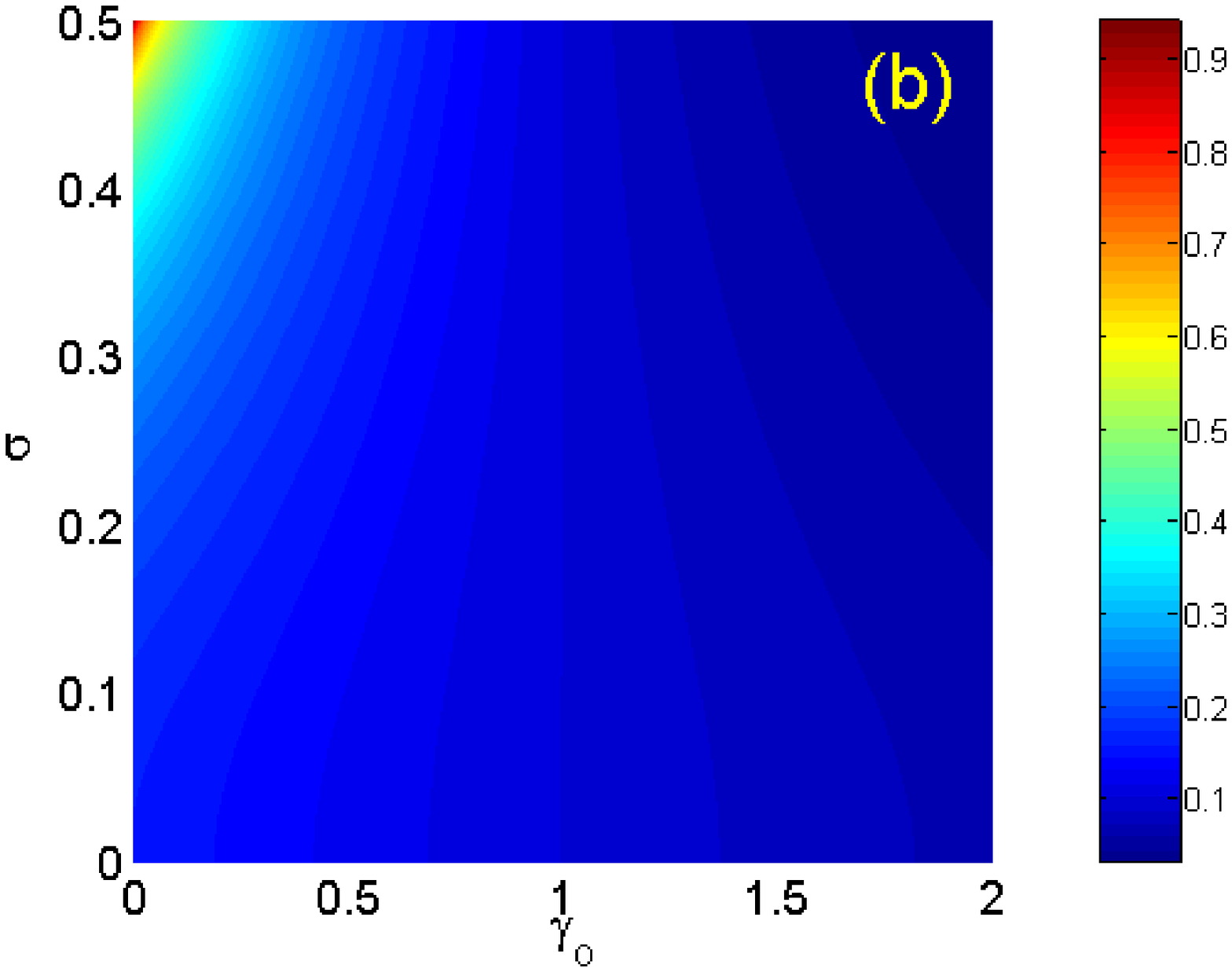} 
  \caption{Plot of the three-body correlation function 
$g_3(\gamma_o,\sigma;\gamma_e)/n^3$ in the $\gamma_o$, $\sigma$-plane. 
Same parameters as in Fig.\ref{fig3}.
} \label{fig4}
\end{figure}
%

We have assumed that the system
is optically trapped (no trapping magnetic field). However, the external 
magnetic field $\mathcal{H}$ required to generate a p-wave Feshbach
resonance adds a Zeeman term 
\beqa
\hat{H}_{\text{Zeeman}}=-g\mu\mathcal{H}\hat{S}^z
\eeqa
to (\ref{Fermi Hamiltonian}) and (\ref{LLH Hamiltonian}), where
$\hat{S}^z=\sum_{j=1}^N \hat{S}_j^z$ is the total spin z-component
operator. 
Such term only shifts the ground state energy without affecting 
the wavefunction. The stability of the gas for an arbitrary polarization 
can be analyzed by means of the same variational
method used in \cite{Gir06b}, with a trial function 
$\psi_{\text{space}}\psi_{\text{spin}}$ where now $\psi_{\text{spin}}$ 
is taken to be the lowest state of the antiferromagnetic Heisenberg model
with total spin $S$, with Hamiltonian 
$\sum_{j=1}^{N-1}\hat{\mathbf{S}}_j\cdot\hat{\mathbf{S}}_{j+1}
+\hat{\mathbf{S}}_N\cdot\hat{\mathbf{S}}_1$.
It has been determined exactly by Griffiths \cite{Gri64}. Denoting its
ground state energy by $E_{0,\text{Heis}}(S)$, one has 
\beqa
\langle\hat{\mathbf{S}}_j\cdot\hat{\mathbf{S}}_{j+1}\rangle_0
=N^{-1}E_{0,\text{Heis}}(S),
\eeqa
 which can be substituted into
Eq. (\ref{LLH Hamiltonian}) to obtain a LL Hamiltonian with LL coupling
constant
\beqa
\label{czeeman}
c=\frac{3c_o+c_e}{4}+(c_o-c_e)\frac{E_{0,\text{Heis}}(S)}{N}.
\eeqa
Note that always $c\ge 0$ implying repulsive interactions and therefore
excluding the possibility of pairing investigated in a 1D 
$\delta$-interacting Fermi gas in an external magnetic field 
\cite{GBLB07}. 

Actually, the magnetization can be changed 
to an arbitrary value using the microwave technique and remains constant 
in a given experiment.  Indeed, experiments often employ the lowest Zeeman 
sublevel of the atom, say $|F=9/2, m_F=-9/2\ra$ for $^{40}$K, in combination with the next
lowest state ($m_F=-7/2$); for which spin-changing collisions are
energetically disfavoured. When tuning the p-wave interactions, the atoms 
are prepared in the $|F=9/2,m_F=-7/2\ra$ state from which depolarization 
into $m_F=-9/2, -5/2$ also is suppressed by the second order 
Zeeman effect \cite{Koehl,Koehl2}.

The even-wave interactions are constrained by the tranverse frequency of the waveguide and linear density 
and therefore $\gamma_e$ is considered as a parameter. The linear densities accessible in current experiments vary within the range $n=0.2-2$ $\mu$m$^{-1}$, the transversal frequency $\om_{\perp}\sim100$kHz and the background 3D even-wave scattering length of $^{40}$K is $a_{bg}=104a_{0}$. Using Olshanii's relation $a_{1D}^e=-\frac{a_{\perp}^2}{2a_{bg}}(1-1.4603\frac{a_{bg}}{a_{\perp}})$ where $a_{\perp}=\sqrt{2\hbar/m\om_{\perp}}$, 
it follows that the typical values of $\gamma_e\approx0.1-10$.
In Fig. \ref{fig3} and \ref{fig4} $g_2$, which determines the photoassociation
rate, and $g_3$, which determines the rate of three-body recombination, are
shown as a function of the magnetization $\sigma=S/N$ and $\gamma_o$ for different values of $\gamma_e$. 
It is clear from Fig. 4 that states of 
lower magnetization are more stable and for larger values of $\gamma_e$, 
that is, stronger s-wave interactions, the region of stability increases.

\section{Conclusions} 

Using the mapping between a 1D spinor Fermi gas and the 
Lieb-Liniger-Heisenberg model, the two and three-body correlation functions
$g_2$ and $g_3$ have been calculated. It is found that $g_3$ is small enough
in a wide area of the $\gamma_o,\gamma_e$-plane, encompassing both ferromagnetic
and antiferromagnetic phases, to ensure stability of this
system against three-body recombination over experimental lifetimes.
Due to the different decay scales of $g_2$ and $g_3$ it may be possible to perform 
photoassociation experiments in a range of p-wave interactions 
($\gamma_o\simeq 2$ in Figs. \ref{fig3} and \ref{fig4})
where $g_2$ is significant and the gas stable because of a negligible $g_3$. 
The limiting case of the fermionic Tonks-Girardeau gas, 
falls however within the region of instability.
Finally, by means of a variational ansatz the results are extended 
for an arbitrary spin polarization. 

\begin{acknowledgments}
This research was supported by U.S. Office of Naval Research grants 
N00014-03-1-0427 and N00014-06-1-0455 through subcontracts from the University 
of Southern California and by Ministerio de Educaci\'on y Ciencia
(FIS2006-10268-C03-01) and UPV-EHU (00039.310-15968/2004). 
A.C. acknowledges financial support by the Basque Government (BFI04.479). 
The authors further acknowledge discussions with 
M. Koehl, M.A. Cazalilla and M. Raizen.

\end{acknowledgments}
\end{document}